\documentclass[11pt,a4paper]{article}
\pdfoutput=1   
\usepackage{amsmath,amsfonts,amssymb,amsbsy}
\usepackage{mathrsfs,latexsym}
\usepackage{graphicx}
\usepackage{color}
\usepackage{booktabs}
\usepackage{multirow}
\usepackage{multicol}
\usepackage{subfig}
\usepackage{alpha}
\usepackage{macros_alpha}
\renewcommand{\strange}{\mathrm{strange}}
\usebiblio{jbulava2,latticen}

\begin{document}

\preprintno{%
TCD 14-04\\
DESY 14-049\\
}

\title{%
Precision lattice QCD computation of the  $B^*B\pi$ coupling
}

\collaboration{\includegraphics[width=2.8cm]{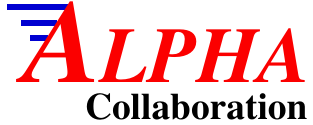}}

\author[desy]{Fabio~Bernardoni}
\author[trin]{John~Bulava}
\author[desy]{Michael~Donnellan}
\author[desy]{Rainer~Sommer}

\address[desy]{NIC @ DESY, Platanenallee~6, 15738~Zeuthen, Germany}
\address[trin]{School~of~Mathematics, Trinity~College, Dublin~2, Ireland}

\begin{abstract}
The static $B^{*}B\pi$ coupling, $\hat{g}_\chi$, 
a low energy constant in the leading order heavy meson chiral Lagrangian,
is determined using $\nf = 2$ lattice QCD. 
We use CLS ensembles with lattice spacings and pion masses down to
$a = 0.05\mathrm{fm}$ and $m_{\pi}=270\mathrm{MeV}$, and perform combined 
continuum and chiral extrapolations of our results which have
a much better accuracy than previous numbers in the literature.
As a by-product, we determine the coupling between the first radial excitations 
in the $B$ and $B^{*}$ channels ($\hat{g}_{22}$). Accounting for all uncertainties,
which are dominated by the chiral extrapolation, we obtain  $\hat{g}_\chi = 0.492(29)$, while 
$\hat{g}_{22}$ is somewhat smaller. The comparison to a precise 
quenched computation suggests that there is
little influence by the sea quarks and $\hat{g}_\chi$ will not change
much when
a dynamical strange quark is included.
\end{abstract}

\begin{keyword}
Lattice QCD \sep Heavy Quark Effective Theory \sep Bottom quarks \sep Heavy Meson chiral  Effective Theory
\PACS{%
12.38.Gc\sep 
12.39.Hg\sep 
14.65.Fy\sep 
12.15.Ff}    
\end{keyword}

\maketitle

\section{Introduction}\label{sec:intro}

Low energy QCD is described by an effective theory based on spontaneously 
broken global $SU(\nf)_L\times SU(\nf)_R$ chiral symmetry, where $\nf$ is the 
number of light quark flavors. At the same time, a low energy expansion of 
hadrons with a single heavy quark with mass $m_h$ exists and  
is known as Heavy Quark Effective Theory
\cite{stat:eichhill1,stat:symm1,stat:symm3,Eichten:1990vp}.
These chiral and heavy quark expansions may be combined to construct 
effective theories for
the low-energy dynamics of hadrons containing a single heavy 
quark~\cite{Burdman:1992gh,Yan:1992gz,Wise:1992hn}. 

The theory that describes mesons is called Heavy Meson Chiral Perturbation Theory 
(HM$\chi$PT) and contains
a single additional leading low energy constant with respect to standard $\chi$PT. 
This additional low energy 
constant, $\hat{g}_\chi$, describes the coupling of heavy mesons to  
pseudo-Goldstone bosons in the chiral ($m_{\pi}^2 \rightarrow 0$) and static 
($m_h \rightarrow \infty$) limits.

The coupling $\hat{g}_\chi$ is relevant for the computation
of B-physics matrix elements from lattice QCD, exemplified by the ALPHA collaboration HQET 
program~\cite{Bernardoni:2013xba,Blossier:2010mk,Blossier:2010vz,Blossier:2010jk,Bernardoni:2014fva}; 
it enters in chiral extrapolations of hadronic parameters needed for 
heavy flavor phenomenology, such as the B-meson decay constant and 
the B-meson semi-leptonic decay form factors. This coupling is
also referred to as the $B^{*}B\pi$ coupling, where the pseudoscalar and vector 
static-light mesons are denoted $B$ and $B^{*}$. Note that
the static ($m_h \rightarrow \infty$) limit is implied.
  
One way to determine $\hat{g}_\chi$ is through phenomenological 
fits to experimental data. A determination from $D^{*}\rightarrow D\pi$
decays~\cite{Anastassov:2001cw}  yields a value of $\hat{g} = 0.61(6)$. 
However, this extraction is affected by $\mathrm{O}(1/m_c)$ and 
$\mathrm{O}(m_{\pi}^2)$ errors, where especially the first ones are hard to 
estimate.
Unfortunately, the process $B^{*}\rightarrow B\pi$ is kinematically forbidden, 
complicating the estimation of the $\mathrm{O}(1/m_h)$ errors from experimental 
results. For a recent review of
results, including also quark model and QCD sum rules calculations, see Ref.~\cite{ElBennich:2010ha}.

In this work we employ a different 
approach. Using lattice QCD simulations, we calculate a 
matrix element of the (light-light) axial current in QCD which is equivalent to $\hat{g}_\chi$ in leading order HM$\chi$PT.
Namely, we compute (ignoring renormalization and improvement in this introduction)
\begin{align} 
\hat{g} = \frac{1}{2}\langle B^0(\textbf{0}) | \hat{A}_k(0) | B^{*+}_{k}(\textbf{0})\rangle, \qquad \hat{A}_{\mu}(x) = {\psibar_d}(x)\gamma_{\mu}\gamma_5\psi_{u}(x),  
\end{align}
where $\psi_d$($\psi_u$) annihilates a down(up) quark and the index 
$k=1,\,2,\,3$ is not summed over. We 
use the finite volume normalization of states 
$\langle B^0(\textbf{p}) | B^{0}(\textbf{p}) \rangle = \langle B^*(\textbf{p}) | B^{*}(\textbf{p}) \rangle = 2L^{3}=2V$, where $L$ is the linear size of the
simulated torus. 
We work directly in the static limit for the heavy quark, but at finite 
light quark mass. Therefore $\hat{g}_\chi$ is eventually obtained 
by an extrapolation of our 
results for $\hat{g}$ to the zero light quark mass (chiral) limit as well as the 
$a\rightarrow 0$ (continuum) limit, where $a$ is the lattice spacing of our 
simulations. 

There have been previous determinations of $\hat{g}_\chi$ using lattice QCD with $\nf = 0,2,3$ dynamical light quark flavors directly in the static limit~\cite{deDivitiis:1998kj,Ohki:2008py,Becirevic:2009yb,Detmold:2012ge}, as well as determinations at the charm~\cite{Becirevic:2012pf} and bottom~\cite{Flynn:2013kwa} 
points. However, the lattice spacing dependence of 
this quantity has not yet been thoroughly investigated. In this work we 
perform a continuum extrapolation in both the $\nf=0$ and $\nf = 2$ theories 
and find small lattice spacing effects for our $\rmO(a)$ improved discretisation. 

\begin{figure}
	\centering
	\includegraphics[width=0.8\textwidth]{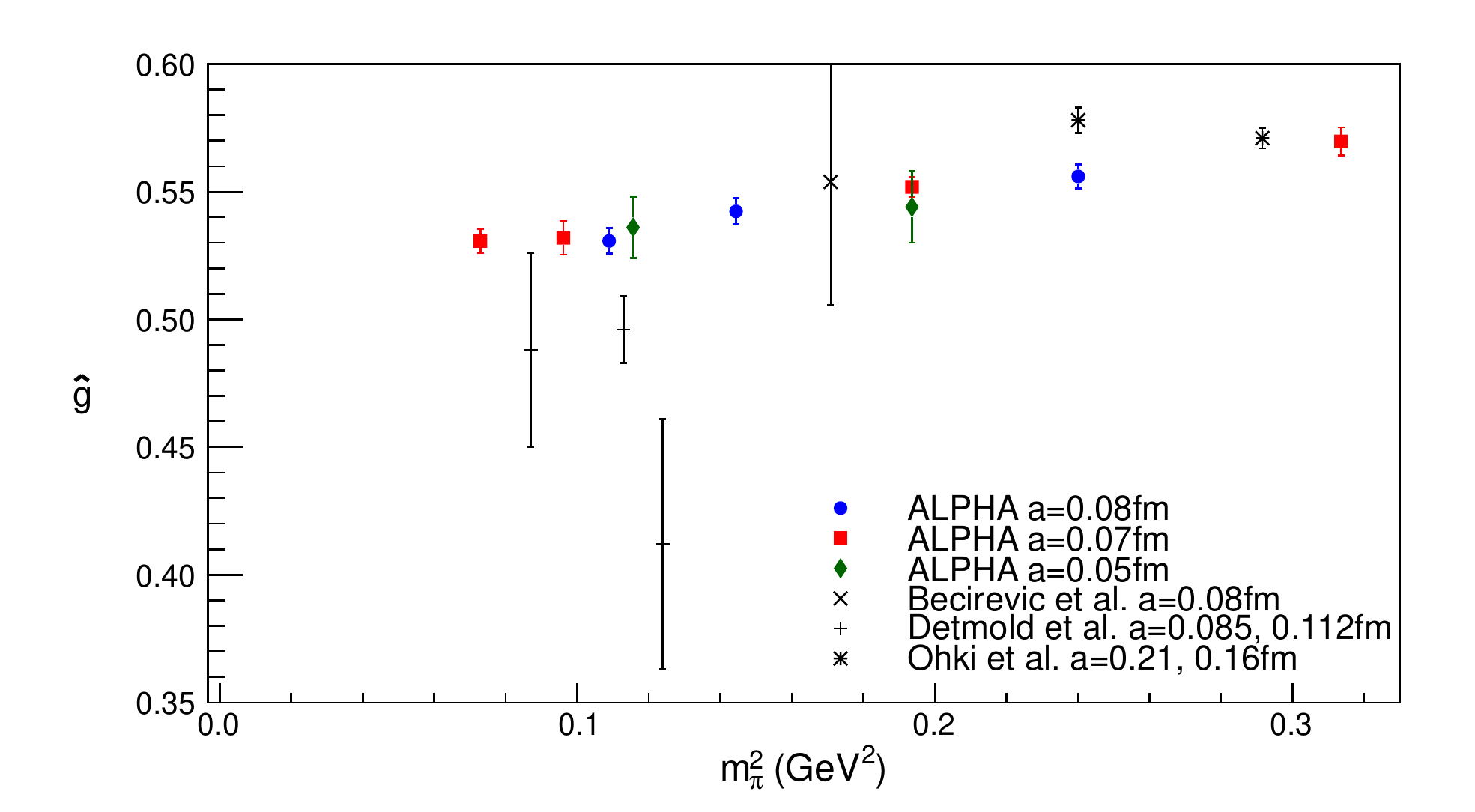} 
	\caption{\label{fig:summary} A summary of unquenched lattice QCD results for $\hat{g}$. The results from the three lattice spacings used in this work are labeled `ALPHA'. Additionally
there are results of Ohki et al.~\protect\cite{Ohki:2008py},
Becirevic et al.~\protect\cite{Becirevic:2009yb} and 
Detmold et al.~\protect\cite{Detmold:2012ge}. 
For Ref.~\protect\cite{Detmold:2012ge}, which employs $\nf = 2+1$ dynamical flavors, we take the results for a single level of link smearing in the static action. } 
\end{figure}

In Fig.~\ref{fig:summary} we compare our results (before extrapolations)
to recent lattice results 
from Refs.~\cite{Ohki:2008py,Becirevic:2009yb,Detmold:2012ge}. We observe that a 
new quality is reached, reducing previous uncertainties by an order of 
magnitude in the region of interest, namely at small 
pion masses\footnote{We do not have access to the numerical results of 
	Ref.~\cite{Flynn:2013kwa} but their Fig.~2 shows errors similar to 
Refs.~\cite{Becirevic:2009yb,Detmold:2012ge}.}.
This is achieved by both improved techniques~\cite{Bulava:2011yz} and good statistics.

Additionally, we quote results for the matrix element of the first radial 
excitations. To this end we define  
\begin{align}
\hat{g}_{mn}(y,a) = \frac{1}{2}\langle B^{0}(\textbf{0}), m | \hat{A}_k(0) | B^{*+}_k(\textbf{0}), n \rangle   
\end{align}
where $m,n=1$ are the ground states of $B$ and $B^{*}$ mesons while $m,n > 1$ 
refer to their excitations. Apart from our main object of study, 
$\hat{g}_{11} = \hat{g}$, we quote rough numbers for 
$\hat{g}_{22}$. Preliminary results for $\hat{g}_{11}$ and $\hat{g}_{22}$ 
have appeared in Ref.~\cite{Blossier:2013qma} (together with results for 
$\hat{g}_{12}$) and a preliminary account of our present work can be found in 
Ref.~\cite{Bulava:2010ej}. The variable $y$ is proportional to the 
square of the pion mass and will be defined when we discuss the 
chiral and continuum limit to arrive at 
$\hat{g}_\chi\equiv \hat{g}_{11}(0,0)$ and $\hat{g}_{22}^\chi\equiv \hat{g}_{22}(0,0)$.

In Sec.~\ref{sec:method} we describe our techniques. In Sec.~\ref{sec:nf0} 
we show results in the quenched approximation, where a continuum limit is 
taken for both $\hat{g}_{11}$ and $\hat{g}_{22}$ at a fixed quark mass 
$m_q \approx m_\strange$. In Sec.~\ref{sec:nf2_results} we discuss the 
$\nf=2$ results, in particular the chiral and continuum extrapolations. Finally we conclude in Sec.~\ref{sec:concl}.

\section{Methodology}\label{sec:method}

Here we describe some details of the lattice calculation of $\hat{g}_{mn}$,
namely the proper definition of the axial current and the technology to obtain 
precise matrix elements from correlation functions including the estimation of systematic errors due to excited state contributions. We also detail our 
stochastic technique utilising translation invariance. 
The ensembles used in the numerical application  
are explained in Secs.~\ref{sec:nf0} and~\ref{sec:nf2_results}.

\subsection{Discretisation and renormalization}

We employ both the HYP1 and HYP2 discretizations of the static quark 
action~\cite{Hasenfratz:2001hp,DellaMorte:2005yc} to mitigate 
the signal-to-noise problem and provide a further check on discretization 
effects. Generally, results from these two 
discretizations are compatible within statistical errors,
but they are also strongly correlated. We will thus show both of them 
in the tables, but only use HYP2 in the detailed analysis.

The light quarks are non-perturbatively $\Oa$ improved Wilson quarks
\cite{impr:pap1,impr:pap3,impr:csw_nf2} and the improved and 
renormalized axial current is
\begin{align}
    (\AR)_k = \za\,(1 + \ba\,a\mq)(A_k + a\ca\partial_k P),
\end{align}
where $A_k$ has exactly the form given before and
$P(x)={\psibar_d}(x)\gamma_5\psi_{u}(x)$ is the appropriate 
pseudoscalar density. For the required values of the bare coupling,
the renormalization constant $\za$ is known 
non-perturbatively  for both $\nf=0$~\cite{Luscher:1996jn} and  $\nf=2$~\cite{DellaMorte:2008xb,Fritzsch:2012wq}
while for the improvement coefficient $\ba$ we use the expansion 
in the bare coupling $g_0^2$ to first order with the one-loop coefficient 
of Ref.~\cite{Sint:1997jx}. For our  zero momentum transfer
matrix element $\hat{g}$, the $\partial_k P$ term vanishes identically;
$\ca$ is not needed.

\subsection{Matrix elements from the GEVP}
The matrix elements $\hat{g}_{mn}$ are accessible in lattice QCD via three-point correlation
functions in Euclidean time, which (from the transfer matrix formalism) have 
the following representation
\begin{align}
\label{e:3pt}
C^{\rm 3pt}_{ij} (t_1, t_2) = a^3\sum_\vecx \langle\, \mathcal{O}_{i}(t_1+t_2)\, (\AR)_k(t_2,\vecx)\, \mathcal{O}_{j}^{k\dagger}(0) 
\,\rangle = \sum_{m,n} \psi_{im} \psi_{jn}^{*} \, 2 \hat{g}_{mn} 
\mathrm{e}^{-E_{m}t_1}
\mathrm{e}^{-E_{n}t_2},
\end{align}
where $\hat{\mathcal{O}}_{i}$ and 
$\hat{\mathcal{O}}^{k}_{j}$ are suitable interpolating fields for the  $B^{0}$ and $B_k^{*+}$  mesons (respectively), 
$\psi_{im} = \langle 0 |\hat{\mathcal{O}}_i   | B,m \rangle = \langle 0 |\hat{\mathcal{O}}^{k}_i | B_k^{*}, m \rangle$ and $E_{m}$ is 
the energy of the $m$th state. In the static limit 
the $B$ and $B^{*}$ energy levels
are the same and furthermore $C^{\rm 3pt}_{ij}$ is independent of $k$ as indicated 
by our notation.

To isolate the desired matrix elements, we also require the two-point 
correlation functions
\begin{align}
C^{\rm 2pt}_{ij}(t) = \langle \mathcal{O}_i(t) \mathcal{O}^{\dagger}_{j}(0) \rangle
 = \sum_m \psi_{im}\psi_{jm}^{*} \mathrm{e}^{-E_{m}t} \,.  
\end{align}
Rather than analyzing $C^{\rm 3pt}(t_1, t_2)$ directly, we employ
\begin{align}
D^{\rm 3pt}_{ij}(t) = \sum_{t_2=0}^{T-1} C^{\rm 3pt}_{ij}(t-t_2, t_2),
\end{align}
where the position of the current insertion is summed 
over~\cite{Maiani:1987by,Bulava:2010ej,Capitani:2012gj}. The use of this 
summed correlation function improves the convergence in $t$ but was proposed in 
Ref.~\cite{Maiani:1987by} for different reasons.

In order to extract the desired matrix elements we choose a set of $N$ 
interpolating operators and form the $N\times N$ correlation matrices 
$C^{\rm 2pt}_{ij}(t)$ and $D^{\rm 3pt}_{ij}(t)$. We then employ solutions of a 
generalized eigenvalue problem (GEVP)~\cite{Michael:1982gb,LuscherWolff,Blossier:2009kd} to accelerate the asymptotic (in $t$) behavior and enable the 
extraction of $\hat{g}_{nn}$ for $n > 1$.  

It has been proven recently that the GEVP may be combined with summed 
insertions~\cite{Bulava:2011yz,Blossier:2013qma} to achieve a further reduction in the contribution from excited states. It was demonstrated that the summed insertion is 
particularly advantageous 
in the extraction of excited state matrix elements when compared to the 
ordinary GEVP. For completeness, we review the main points. We begin by 
solving the following GEVP
\begin{align}
C^{\rm 2pt}(t)v_{n}(t,t_0) = \lambda_n(t,t_0)C^{\rm 2pt}(t_0)v_{n}(t,t_0), 
\end{align}
 where $t_0 \ge t/2$. It can be shown~\cite{Bulava:2011yz} 
 that
 \begin{align}
M^{\rm eff}_n(t,t_0) &\equiv -\frac12\partial_t { ( v_n(t,t_0) , 
[ D^{\rm 3pt}(t)\lambda^{-1}_n(t,t_0) - D^{\rm 3pt}(t_0) ] v_n(t,t_0) )  
\over 
(v_n(t,t_0) , C^{\rm 2pt}(t) v_n(t, t_0))} 
\\
\label{e:asy}
&= \hat{g}_{nn} + 
\mathrm{O}(\mathrm{e}^{-\Delta_{N,n} t}),
\end{align} 
where $\Delta_{N,n} = E_{N+1} - E_n$ and $(.,.)$ denotes an inner product 
over the GEVP indices. The important result is that (asymptotically)
the corrections fall exponentially in $\Delta_{N,n} t$. The large energy gap
$\Delta_{N,n}$, which in our application is above 1~GeV, is a virtue of the 
GEVP and the factor $t$ is due to the  summed insertion. Without summation,
$t$ would be replaced by $\min(t_1,t_2)$, see \eq{e:3pt}. 
We also note  that the GEVP renders
excited state matrix elements accessible. 

In our numerical application, the interpolating fields
$\mathcal{O}_i(x_0) = a^3\sum_\vecx {\psibar_b}(x)\Gamma_i\gamma_5\psi_{d}(x)$
and 
$\mathcal{O}_i^k(x_0) = a^3\sum_\vecx {\psibar_u}(x)\Gamma_i\gamma_k\psi_{b}(x)$ 
are constructed from Gaussian smearing operators 
\begin{align}
	\Gamma_i = (1+ \kappa_\mathrm{G} a^2 \Delta)^{R_i}, \quad i = 1,2, 3, 
\end{align}
where $\Delta$ is the gauge-covariant spatial Laplace operator 
with APE-smeared links.  
The approximate width $r_i \approx 2a\sqrt{\kappa_G R_i}$ is chosen to keep the smearing radii at $r_i\approx 0.2,\,0.3,\,0.7$~fm for each lattice spacing. 
More details about the construction of 
these wave-functions can be found in Refs.~\cite{Blossier:2010vz,Bernardoni:2013xba}.  

From the correlation functions we construct $M^{\rm eff}_n(t) \equiv M^{\rm eff}_n(t,t-a)$ and examine the large $t$ behaviour. Beginning with $t \approx r_0 \approx 0.5\mathrm{fm}$, we increase $t$ 
 until the asymptotic corrections due to excited states are small enough so that 
\begin{align}
\label{e:platcrit}
|M^{\rm eff}_n(t) - M^{\rm eff}_n(t - \delta t)| 
<  \sigma(t)
\end{align}
where $\delta t = \frac{1}{\Delta_{N,n}}$ and $\sigma(t)$ is the statistical ($1\sigma$) error 
on $M^{\rm eff}_n(t)$. We call the first $t$ at which this condition is satisfied $t_{\rm min}$. Under the assumption that the asymptotic
decay \eq{e:asy} has roughly set in at $t_{\rm min}$, our requirement
of \eq{e:platcrit} means that statistical errors exceed systematic ones
by a factor $e-1 \approx 2$ at $t=t_{\rm min}$. 
We then define our estimate of $\hat{g}_{nn}$ as the weighted average of $M^{\rm eff}(t)$ over 
the range $[t_{\rm min}, t_{\rm max}]$ with $t_{\rm max}$ chosen
to avoid points with excessive statistical errors. The estimate for the 
statistical error on $M^{\rm eff}(t)$ will be discussed in future sections, in particular for the $\nf =2$ results where autocorrelations must be treated with care.  

\subsection{Use of random sources}
In order to reduce statistical fluctuations, we use full translation invariance 
everywhere. We achieve this by a stochastic estimation of one of the 
spatial sums, using \cite{lat94:rainer} a random $U(1)$ source on each time-slice of the lattice
(`time-dilution'~\cite{Foley:Practical}) and a `sequential inversion' (\eq{e:dirac2} below) for the insertion of the axial current. 

The explicit expression for the two-point function 
is
\begin{align}
	&C^{\rm 2pt}_{ij}(y_0-t_0) = 
	a^7  \sum_{\vecy,x} \langle\,\eta_{r}^\dagger(x;t_0) \Gamma_j\gamma_5\,
	         S_b(x,y) \Gamma_i\gamma_5 \Phi_r(y;t_0) \, \rangle  \,,
	\\
	& [\mathcal{D}\,\Phi_{r}](y;t_0) = \eta_{r}(y;t_0) \,,
	\label{e:dirac1}
\end{align}
where $\eta_{r}(y;t_0)\propto \delta(y_0-t_0) \in U(1)$ is a random 
U(1) field on timeslice $t_0$ and vanishes otherwise, 
$S_b(x,y)$ is the easily computed static quark propagator of the 
b-quark and $\mathcal{D}$ is the Dirac operator of the light quarks. 
The subscript $r$ enumerates the random source fields.
Similarly, the three-point functions with summed current insertions read
\begin{align}
	&D^{\rm 3pt}_{ij}(y_0-t_0)  =
	\za\,(1 + \ba\,a\mq)\, a^7 \sum_{\vecy,x}\langle\,\eta_{r}^\dagger(x;t_0) \Gamma_j\gamma_k\,
	          S_b(x,y) \Gamma_i\gamma_5 \tilde\Phi_{r,k}(y;t_0) \, \rangle \,,
	\\
	&    [\mathcal{D}\,\tilde\Phi_{r,k}](y;t_0) = \gamma_5\gamma_k \Phi_{r}(y;t_0) \,.
	\label{e:dirac2}
\end{align}
In the numerical application, we average over 
$r=1\ldots N_r$, all values $0\leq t_0 \leq T-a$ and over
 $k=1,2,3$. Hence,  \eq{e:dirac1} and \eq{e:dirac2}
 add up to $N_{r} \times T/a \times 4$  Dirac equations which
 need to be solved on each gauge configuration. 
 The ensemble average $\langle \,.\,\rangle$
indicates an average over the random fields $\eta_r$ as well as the gauge fields.

\section{Results}\label{sec:res}

We now discuss numerical results for 
$\nf=0$ and $\nf=2$. In both cases we employ the usual 
periodic and anti-periodic temporal boundary 
conditions for the gauge and fermion fields, respectively. 

\subsection{$\nf = 0$ continuum limit}\label{sec:nf0}

We first apply the methods discussed in Sec.~\ref{sec:method} to a set of three
ensembles of quenched gauge configurations used previously in the ALPHA 
collaboration HQET program~\cite{Blossier:2010mk}, with the goal of taking the continuum limit of $\hat{g}_{11}$ and $\hat{g}_{22}$. Details of 
the ensembles and measurements are given in Tab.~\ref{tab:qens}. The valence quark mass on each of these ensembles was tuned to reproduce the physical strange quark mass\cite{mbar:pap3}.

\begin{table}
	\centering
\begin{tabular}{|l|l|c|r|l|r|}
	\hline
$\beta$ &  $\kappa$ & $(L^3/a)\times T/a$ & $r_0/a$ & $N_{\rm conf}$ & $N_{r}$ \\
\hline 
6.0219 & 0.133849 & $16^3 \times 32$ & 5.6 & 100 & 200 \\
6.2885 & 0.1349798& $24^3 \times 48$ & 8.4 & 100 & 48 \\
6.4956 & 0.1350299& $32^3 \times 64$  & 11.0& 100 &32\\
	\hline
\end{tabular}
\caption{\label{tab:qens} Details of the quenched ensembles used for the continuum extrapolations of $\hat{g}_{11}$ and $\hat{g}_{22}$. The valence quark mass is chosen to be the strange quark mass~\protect\cite{mbar:pap3} and the values for $r_0/a$ are taken from Ref.~\cite{Guagnelli:1998ud}.}
\end{table}

\begin{figure}
\includegraphics[width=.49\textwidth]{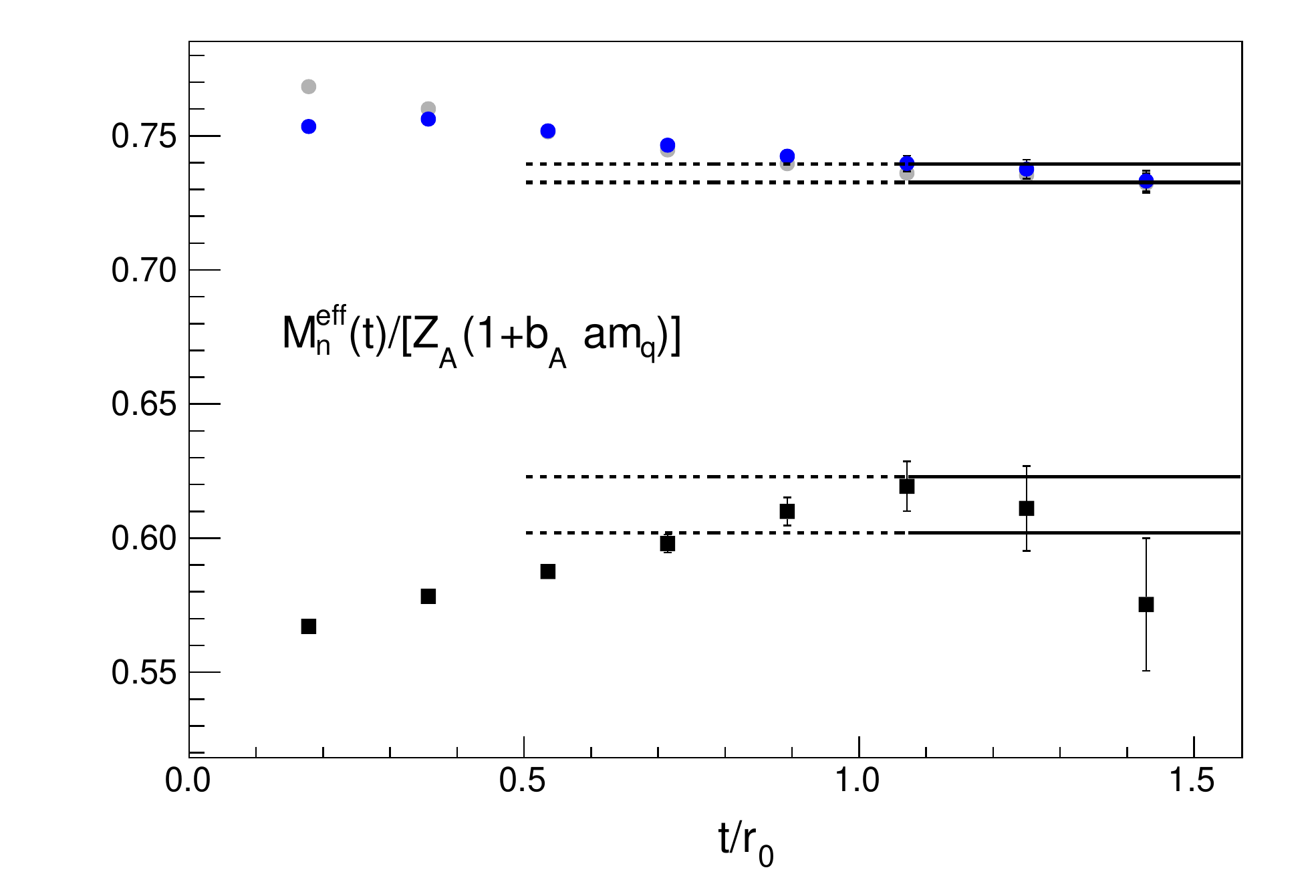}
\includegraphics[width=.49\textwidth]{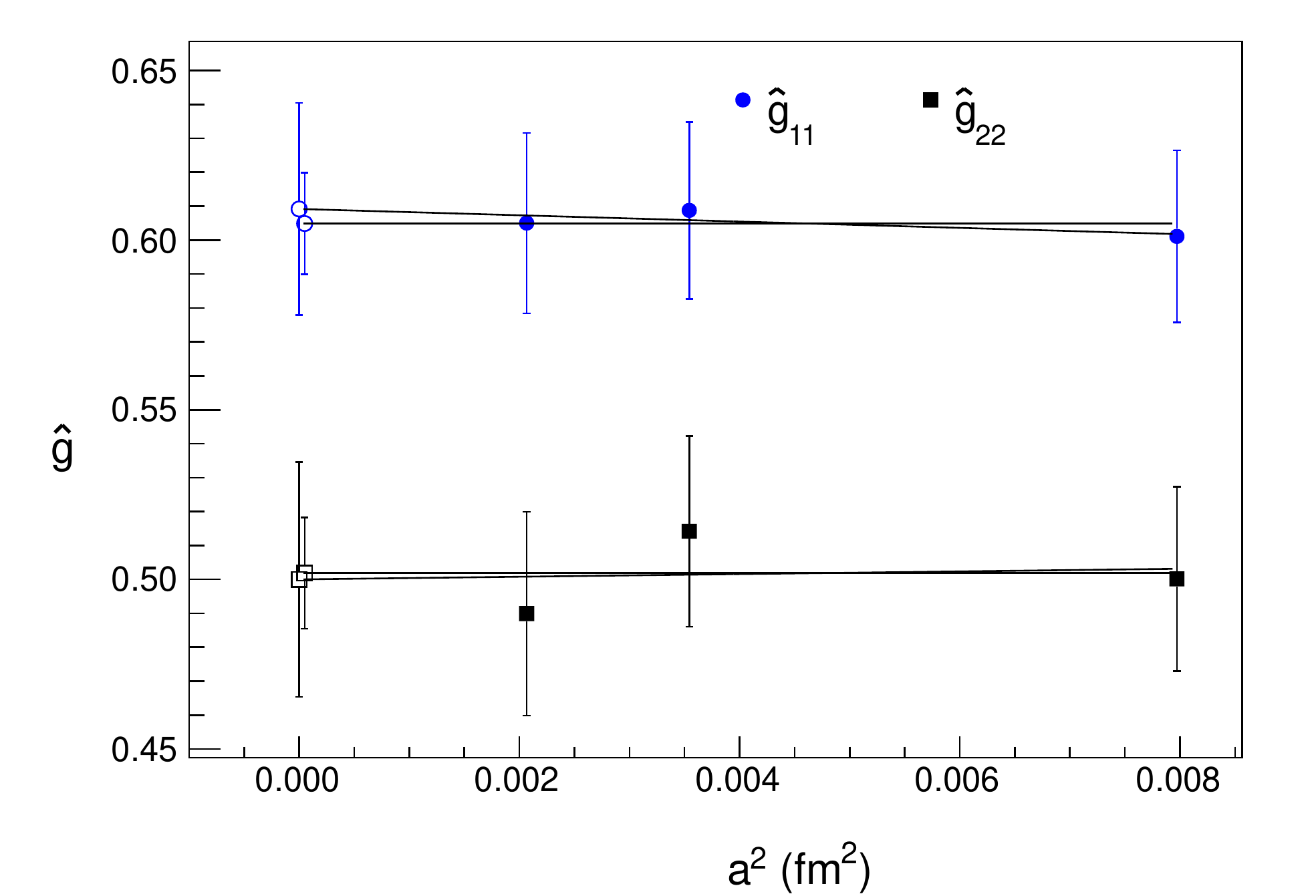}
\caption{\label{fig:nf0_cont} \textbf{Left}: The bare effective matrix elements $M^{\rm eff}_{n}(t)/[\za\,(1 + \ba\,a\mq)]\,,\;n=1,2$ 
for the HYP2 action together with the plateau averages denoted by solid lines. 
The gray points for $n=1$ are the results obtained without the GEVP using the interpolator $\Gamma_3$ with smearing radius $r_3 \approx 0.7\,\mathrm{fm}$, while 
the dotted lines are simply meant to guide the eye. 
\textbf{Right}: The continuum limits 
for $\hat{g}_{11}$ and $\hat{g}_{22}$ for the HYP2 static quark action, taking 
$r_0 = 0.5\mathrm{fm}$.}
\end{figure}

The effective matrix elements $M^{\rm eff}_{n}(t), \: n = 1,2$ for the $L/a = 16$ ensemble are shown together 
with their plateau averages in the left plot of Fig.~\ref{fig:nf0_cont}. 
As autocorrelations play no role here,
the errors are estimated using 100 single-elimination jackknife bins after
first averaging over the $N_{r}$ sources on each 
configuration. The range of the plateau averages is chosen according to the criteria in Sec.~\ref{sec:method}. 

	Finally, we perform continuum extrapolations of the renormalized axial 
	current matrix elements. These extrapolations for $\hat{g}_{11}$ and 
	$\hat{g}_{22}$ are shown in the right plot of Fig.~\ref{fig:nf0_cont} and suggest that cutoff effects are small for these quantities. 
	Simple constant extrapolation of just the  HYP2 action results 
	yields  $\hat{g}_{11}^\strange = 0.605(15) ,\;\hat{g}_{22}^\strange = 0.502(16)$
	while we take 
\begin{align}
\label{e:resqu}
\hat{g}_{11}^\strange = 0.609(31) , \qquad                 
\hat{g}_{22}^\strange = 0.500(35)\;\;\text{for }\nf=0
\end{align}
as our final quenched results, where a linear term in $a^2$ is allowed in 
the fit formula.
It should be noted that our result for $\hat{g}_{11}$ is compatible with the 
previous result of Ref.~\cite{Bulava:2010ej}, but utilizes a more robust 
treatment of the systematic errors due to excited states, namely that of 
Sec.~\ref{sec:method}.

\begin{table}[h!]
	\centering
\begin{tabular}{|l|l|l|l|l|r|r|r|}
	\hline
	ID & $\beta$ & $m_{\pi}(\mathrm{MeV})$ & $(L^3_s\times L_t)/a^4$ & $a(\mathrm{fm})$ & $N_{\rm conf}$ & $\tauexp$ & $N_{r}$ \\
\hline 
A3 & 5.2  & 495  &$32^3\times 64$ & 0.076 & 1004 & 8 & 4  \\
A4 &      & 385  &                &       & 1012 & 8 & 8  \\
A5 &      & 332  &                &       & 500  & 6 &4  \\
\hline
E4 & 5.3  & 577  &$32^3\times 64$ & 0.066 & 157  & 17 &48 \\
E5 &      & 440  &                &       & 1000 & 8  &4  \\
F6 &      & 310  &$48^3\times 96$ &       & 500  & 17 &4         \\
F7 &      & 270  &                &       & 461  & 17 &4   \\
\hline 
N5 & 5.5  & 440  & $48^3\times96$ & 0.048 & 476  & 110 &2   \\
N6 &      & 340  &                &       & 400  & 25 &2   \\
\hline
\end{tabular}
\caption{\label{tab:nf2ens} Details of the $\nf = 2$ ensembles. The 
pseudoscalar meson masses and lattice spacings are taken from 
Ref.~\protect\cite{Fritzsch:2012wq} and $N_r$ is as defined in Sec.~\protect\ref{sec:method}. We also list our estimate of the exponential 
autocorrelation time $\tauexp$ in units of the separation
between configurations.}
\end{table}

\subsection{$\nf = 2$ results}\label{sec:nf2_results}

We next use ensembles of 
the Coordinated Lattice 
Simulations (CLS) community effort.
Details are tabulated in Tab.~\ref{tab:nf2ens}. 

While we follow the same procedure to calculate the bare matrix elements, 
the large autocorrelations present in HMC simulations with periodic 
 boundary conditions must be taken into 
account in order to safely estimate the statistical errors. To this 
end, we follow the procedure of Ref.~\cite{Schaefer:2010hu} and attach an 
exponential `tail' to the autocorrelation functions of the matrix 
elements, with a fall-off $\sim \exp(-t_\mathrm{MC} / \tauexp)$,
where $\tauexp$ has been estimated roughly in Ref.~\cite{Schaefer:2010hu}. 

\begin{figure}
\centering
\includegraphics[width=0.9\textwidth]{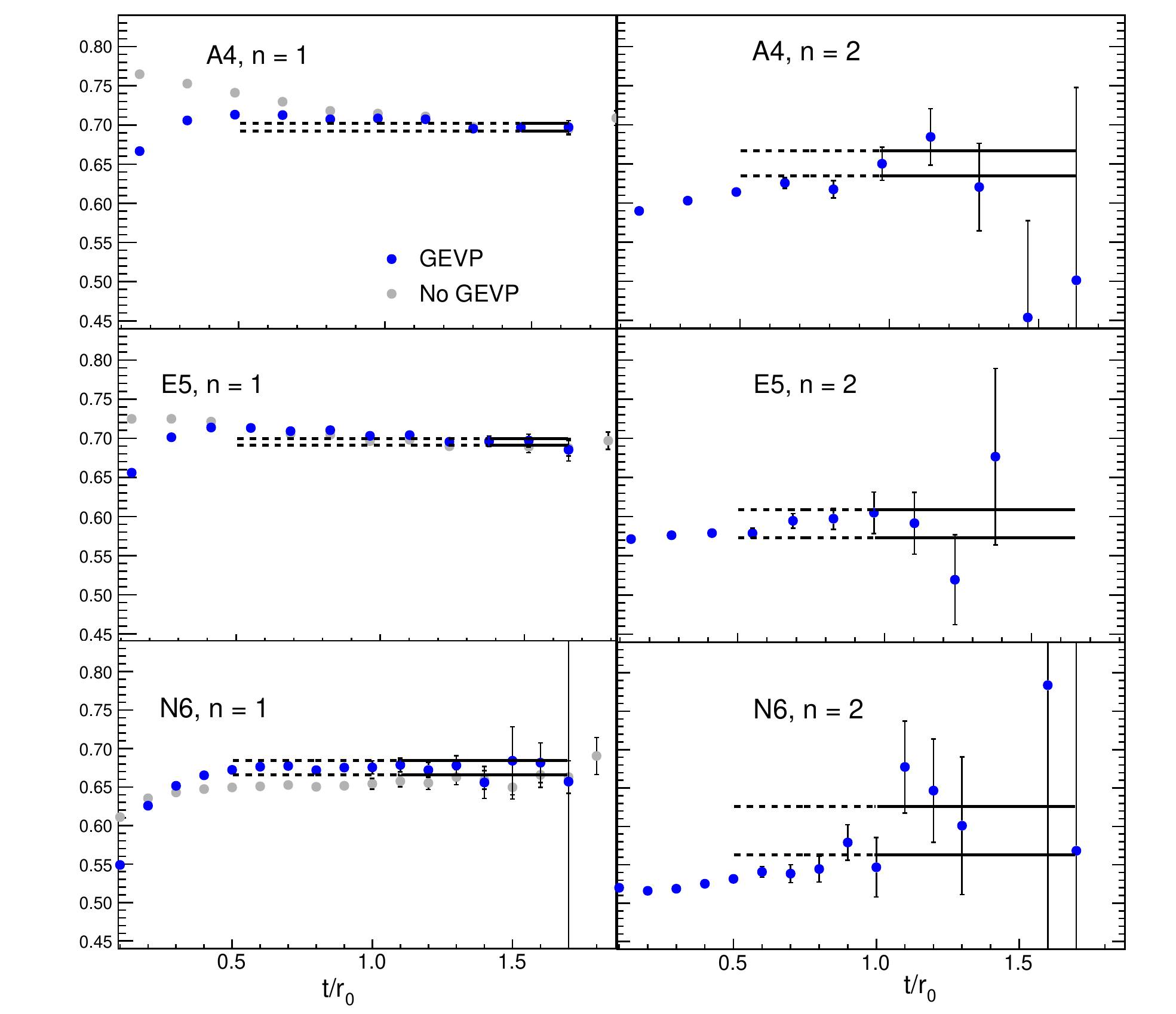}
\caption{\label{fig:nf2_best} The bare effective matrix elements 
for a selection of the $\nf = 2$ ensembles. The left column is 
$M^{\rm eff}_{1}(t)$ while the right is $M^{\rm eff}_{2}(t)$. The left column also 
shows the effective matrix elements obtained without the GEVP represented by 
the gray points as in Fig.~\ref{fig:nf0_cont}. The solid lines show the 
plateau averages while the dotted lines are meant to guide the eye.}
\end{figure}

A selection of the effective matrix elements 
is shown in Fig.~\ref{fig:nf2_best}. This figure also compares the use of the 
summed insertion with and without the GEVP for $M^{\rm eff}_1(t)$. 
A clear picture for the difference is not easily seen for the few points available. 
However, the figure indicates that while corrections are not 
necessarily smaller with the GEVP at small times $t \approx r_0/2$, the 
approach to the plateau is then accelerated soon after. Indeed, this 
is needed for our criteria \eq{e:platcrit} to apply.

For the excited state matrix element our statistical errors are not small
enough to apply \eq{e:platcrit} to fix the start of the plateaux.
Here we simply inspect the figures and choose a 
fixed $t_\mathrm{min} \approx 0.5\,\mathrm{fm}$ in physical units. 
The large statistical errors seem to
dominate over the systematic errors due to excited state contributions. 
The renormalized matrix elements together with the 
statistical errors estimated using the additional exponential tail are collected in 
Tab.~\ref{tab:nf2_r}.     
\begin{table}
\centering
\begin{tabular}{|l|l|l|l|l|}
\hline	
	Ens. ID& $\hat{g}^{HYP1}_{11}$ & $\hat{g}^{HYP2}_{11}$ & $\hat{g}^{HYP1}_{22}$ & $\hat{g}^{HYP2}_{22}$ \\
	\hline
      A3  &  0.553(5)  &  0.556(5)   &  0.497(15)  &  0.505(14)  \\  
	    A4  &  0.537(6)  &  0.542(5)   &  0.498(14)  &  0.503(15)  \\ 
	    A5d  &  0.528(5)  &  0.531(5)   &  0.516(18)  &  0.527(19)  \\
	\hline
	    E4  &  0.567(5)  &  0.570(6)   &  0.489(13)  &  0.494(14)  \\ 
	    E5g  &  0.543(5)  &  0.546(5)   &  0.460(17)  &  0.464(17)  \\ 
	    F6  &  0.531(7)  &  0.532(7)   &  0.470(17)  &  0.469(16)  \\ 
	    F7  &  0.528(4)  &  0.531(5)   &  0.479(13)  &  0.477(13)  \\ 
	   \hline 
				N5  &  0.541(15)  &  0.544(14)     &  0.477(72)  &  0.477(71)  \\ 
	    N6  &   -         &  0.536(12)     &     -       &  0.465(43)  \\ 
	   \hline 
\end{tabular}
\caption{\label{tab:nf2_r}$\nf=2$ renormalized values for $\hat{g}_{11}$ and 
$\hat{g}_{22}$ for both HYP1 and HYP2 static quark actions. The errors are the combined statistical errors from the bare matrix elements (taking autocorrelations into account) and the renormalization constants. Note that no data for the 
HYP1 action is present for the N6 ensemble.}
\end{table} 

Next, the combined chiral and continuum extrapolation is performed. 
Let us first concentrate on the ground state matrix element, which yields 
$\hat{g}_\chi$. We parameterize the quark mass dependence 
by the pion mass (as in Ref.~\cite{Bernardoni:2013xba}) through the variable\footnote{Our normalization of the 
pion decay constant is such that it is
$f_\pi\simeq 130\mathrm{MeV}$ in the chiral limit.}
$y={m^{2}_{\pi}}/{(8\pi^2f_{\pi}^2})$.

As already evident from Fig.~\ref{fig:summary}, the data is
rather linear in $\mpi^2$ (or $y$), 
while chiral perturbation theory predicts a significant 
logarithmic modification \cite{Detmold:2011rb,Detmold:2012ge}. 
We therefore perform two extrapolations to the chiral limit. Namely 
we fit to the two forms
\begin{align}
\label{e:linfit}
\hat{g}^\mathrm{lin}_{11}(y, a) &= \hat{g}_\chi + By + C a^2 ,  \\ 
\label{e:chifit}
\hat{g}^{NLO}_{11}(y, a) &= \hat{g}_\chi 
    \big[1 - (1 + 2(\hat{g}_\chi)^2)y\log y\big] + By + 
Ca^2,  
\end{align}
where $B$, $C$, and the desired $\hat{g}_\chi$ are fit parameters.
In both forms, the $C a^2$ term can probe cutoff effects and
we consider terms of order $y a^2$ as too small to be relevant, as we do with 
$a^3$ and $y^2$. In both fits, the results for $C$ are compatible with zero.
As our central values we take the fit 
results for $\hat{g}_\chi$ with $C$ set to zero. 
	\begin{figure}
\includegraphics[width=\textwidth]{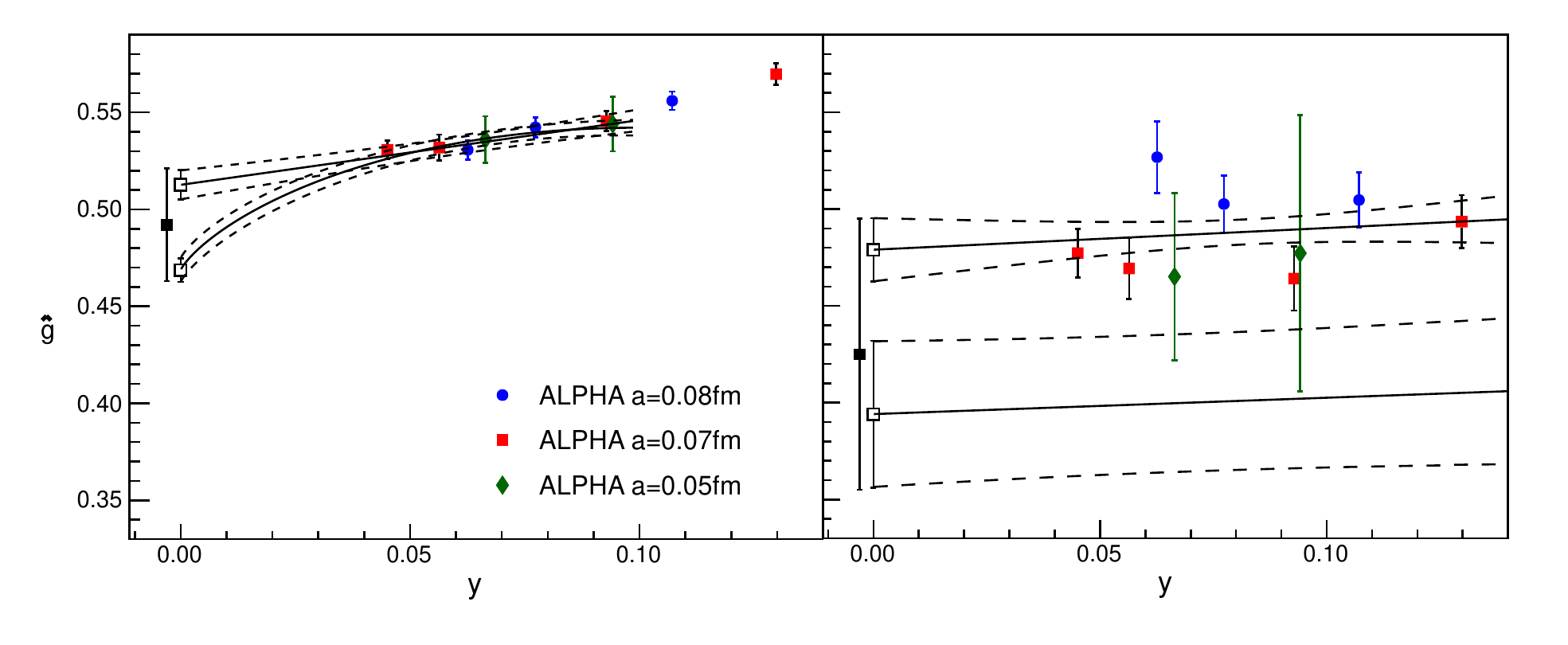}
\caption{\label{fig:nf2_fits} A summary of $\hat{g}_{11}$ 
(left) and $\hat{g}_{22}$ (right) together with the 
extrapolated values and final results explained in the text. 
For $\hat{g}_{11}$, both linear and NLO chiral extrapolations are shown, 
while for $\hat{g}_{22}$ we only perform 
a linear extrapolation. In the case of the excited state 
matrix element both the extrapolation with $C$ as fit parameter (lower curve) and the one with $C=0$ are shown.} 
\end{figure} 
The linear fit, \eq{e:linfit}, then yields 
$\hat{g}_\chi=0.513(8)$ while the correct {\em asymptotic} form,
\eq{e:chifit}, extrapolates further down to
$\hat{g}_\chi=0.469(7)$. We combine these numbers to our central result
\begin{align}
    \label{e:chival}
		\hat{g}_\chi = 0.492(29)\,.
\end{align}
The error is by far dominated by the difference of 
the two chiral extrapolations. 
We have chosen a range which encompasses the linear and the NLO
extrapolation {\em and} their errors. 
Allowing for non-vanishing $C$ changes rather little concerning
this result.

Of course the situation is far from perfect: the theoretically
well motivated functional form is not verified by the 
data; a linear dependence fits somewhat better. However, in
the end we are interested in the extrapolated value 
and it seems very safe to assume that it lies in the range
\eq{e:chival}. Indeed, if NLO chiral behaviour sets in at masses which
are below the ones in Fig.~\ref{fig:nf2_fits}, the downward bend
will happen later and the result will be in between the two values
shown in the figure and used to form \eq{e:chival}.

For $\hat{g}_{22}$ the functional form including
chiral log's is not known. Also the data are much less precise.  
We thus perform a simple linear extrapolation both with and without an 
$a^2$ term. They are shown on the right side of
Fig.~\ref{fig:nf2_fits}. A range covering both results is
\begin{align}
		\hat{g}_{22} =  0.425(70).
\end{align}

\section{Conclusions}\label{sec:concl}

In this paper we have presented a precise $\nf =2$ determination of 
$\hat{g}_\chi$, the leading low energy constant appearing in HM$\chi$PT 
parametrizing the coupling of heavy-light mesons to pions. We have calculated the bare matrix elements using solutions of the GEVP together with the summed 
insertion technique, resulting in a precision which exceeds previous ones by an order of magnitude.
We renormalized these matrix elements non-perturbatively.  

We have taken 
the continuum and chiral limits assuming both a phenomenological
linear behaviour in the square of the pion mass and
next-to-leading-order continuum HM$\chi$PT. Two discretizations of the
static quark action serve as a further check on lattice spacing effects.
These two discretizations give statistically compatible results for 
all quantities. Our central result is
\eq{e:chival}. This value, $\hat{g}_\chi = 0.49(3)$, improves
in accuracy compared to previous estimates: 
$\hat{g}^\mathrm{quenched} = 0.42(4)(8)$ \cite{deDivitiis:1998kj}, 
$\hat{g}_\chi=0.52(1)(3)(3)$ \cite{Ohki:2008py},
$\hat{g}_\chi=0.44(3)(^7_0)$ \cite{Becirevic:2009yb}, 
$\hat{g}_\chi= 0.45(5)(2)$  \cite{Detmold:2012ge}, 
$\hat{g}_\chi^\mathrm{charm}=0.53(3)(3)$ \cite{Becirevic:2012pf} and 
$\hat{g}_\chi^\mathrm{bottom} = 0.57(5)(6)$ \cite{Flynn:2013kwa}.
Within the cited overall uncertainties all previous numbers are in 
agreement with our more precise value. In fact the agreement is better 
than one might have expected given that some numbers come from
extrapolations from rather large pion masses and lattice spacings.

As discussed in Sec.~\ref{sec:method}, we have treated the systematic errors 
due to excited states in a conservative manner. Similarly, we have also safely estimated the statistical 
errors by including tails in the autocorrelation functions. 
At our finest 
lattice spacing, these 
are significant, see Sec.~\ref{sec:nf2_results}.
However, in the end, the dominating uncertainty comes from the fact
that the data does not appear to be at such small pion masses
where NLO chiral behaviour can be seen. Instead an approximately
linear behaviour in $\mpi^2$ prevails down to $\mpi=270$~MeV. 
We thus take a final range which also covers the result of 
a simple linear extrapolation.

In comparison to the quenched result, we have to take into 
account that \eq{e:resqu} is for a light quark mass set to the 
strange mass. This corresponds to 
$y\approx0.2$, outside the range of Fig.~\ref{fig:nf2_fits}. The figure then
suggests that $\nf=2$ and the quenched number agree within
 at least 5\% precision. We 
do not see any sea quark effects at the strange mass. It thus appears safe to use
\eq{e:chival} with its more than 5\% error 
also for the three (or more) flavor theory. 

Despite our limited control over the chiral limit, 
our determination of $\hat{g}_\chi$ is precise enough to help the chiral 
extrapolation of many quantities of phenomenological interest in heavy meson 
physics. 
For example, this result is used broadly in the ALPHA collaboration HQET program.

\vskip 0.3cm

\noindent
{\bf Acknowledgements.}
We thank Hubert Simma, David Lin, and Gilberto Colangelo for useful 
discussions, Benoit Blossier and Antoine G\'{e}rardin for providing us 
with data on the N6 ensemble from Ref.~\cite{Blossier:2013qma}, and Patrick Fritzsch for useful comments on an earlier version of this manuscript. 
This work is supported by the Deutsche Forschungsgemeinschaft in the
SFB/TR 09 and by the European community through EU Contract No.
MRTN-CT-2006-035482, ``FLAVIAnet''. We are grateful to NIC and to the
Norddeutsche Rechnerverbund for allocating computing resources to this
project. Some of the correlation function measurements were performed
on the PAX cluster at DESY, Zeuthen.


\end{document}